\RequirePackage{fix-cm}
\documentclass[twocolumn]{svjour3}          
\smartqed  
\usepackage{graphicx,color,soul}
\usepackage[normalem]{ulem}
\journalname{JOSC}
\begin{document}

\title{High-T$_c$ cuprates: a story of two electron subsystems\thanks{Supported by the European Research Council under ERC Consolidator Grant No. 725521, project CeNIKS co-financed by the Croatian Government and the European Union through the European Regional Development Fund-Competitiveness and Cohesion Operational Programme (Grant No. KK.01.1.1.02.0013) and by the Croatian Science Foundation under Project IP-2018-01-7828.}
}


\author{N.~Bari\v si\'c         \and
        D.~K.~Sunko
}


\institute{N.~Bari\v si\'c \at
              Department of Physics, Faculty of Science, University of Zagreb, 10000 Zagreb, Croatia \\ \emph{and} \\
              Institute of Solid State Physics, TU Wien, 1040 Vienna, Austria\\
              \email{nbarisic@phy.hr}
           \and
           D.~K.~Sunko \at
              Department of Physics, Faculty of Science, University of Zagreb, 10000 Zagreb, Croatia\\
              \email{dks@phy.hr}
}

\date{\textbf{Dedicated to Prof. Karl Alex M\"uller on the occasion of his 95th birthday.} \\
J Supercond Nov Magn (2022). DOI: 10.1007/s10948-022-06183-y}

\maketitle

\begin{abstract}
A review of the phenomenology and micro\-scopy of cuprate superconductors is presented, with particular attention to universal conductance features, which reveal the existence of two electronic subsystems. The overall electronic system consists of $1+p$ charges, where $p$ is the doping. At low dopings exactly one hole is localized per planar copper-oxygen unit, while upon increasing doping and temperature, the hole is gradually delocalized and becomes itinerant. Remarkably, the itinerant holes exhibit identical Fermi-liquid character across the cuprate phase diagram. This universality enables a simple count of carrier density and yields comprehensive understanding of the key features in the normal and superconducting state. A possible superconducting mechanism is presented, compatible with the key experimental facts. The base of this mechanism is the interaction of fast Fermi-liquid carriers with localized holes. A change in the microscopic nature of chemical bonding in the copper-oxide planes, from ionic to covalent, is invoked to explain the phase diagram of these fascinating compounds.

\end{abstract}

\section{Introduction}

The surprising discovery of high-T$_c$ superconductivity (SC) in cuprates~\cite{Bednorz86} has been one of the major driving forces of research in condensed-matter physics over the past thirty-five years. Its importance, recognized by researchers across the globe, was twofold. First, obviously, the SC transition temperature T$_c$ was higher than some even believed possible. More subtly, the cuprates are oxides, while received wisdom expected oxides to be antiferromagnetic (AF) at most. The search for  SC in perovskite oxides, on which Karl Alex M\"uller embarked in the early seventies, was a lonely one until brought to fruition by an inspired switch from titanates and nickelates to cuprates~\cite{Bednorz87}. The success of this latter idea in collaboration with Georg Bednorz, crowned by a Nobel prize less than a year later, is often hailed as one of the quickest recognitions of a breakthrough in Nobel history. Today, as we honor the instigator of this major development in our field on the occasion of his 95th birthday, it is only proper to remember both his long persistence, and the willingness of his collaborator to join what may well have seemed a forlorn quest already by the eighties.

The reception of this discovery in Zagreb was almost immediate. Three early works on the low-temperature orthorombic/tetragonal (LTO/LTT) transition in LBCO in contrast with the high-temperature tetragonal (HTT)/LTO transition established the active electronic role of the oxygen degree of freedom in cuprate SC~\cite{Barisic87,Barisic88,Barisic90}. In particular, the strong suppression of the SC transition temperature (T$_c$) by the LTT tilt was connected with the splitting of the in--plane O$_x$-O$_y$ site energies, pointing to the crucial importance of ungapped  O$_x$-O$_y$ charge fluctuations for cuprate SC. The oxygen-centered framework established at that time still informs our understanding of the cuprates, presented in this contribution.

\subsection{Superconductivity}

The first outstanding property of superconductivity in cuprates is its persistence to almost half the room temperature, at ambient pressure. Second, the bulk T$_c$ is undiminished in strictly two-dimensional (2D) crystal plains~\cite{Gozar08,Logvenov09,Yu19}, in sharp distinction to SC suppression in thin films of elemental Bardeen-Cooper-Schrieffer (BCS) superconductors, where the pair size ($\sim 250$ lattice spacings in Al~\cite{Pippard53}) requires the sample to be macroscopically three-dimensional (3D). Nevertheless, the SC itself appears rather conventional. For example, the existence of Cooper pairs, made of standard quasiparticles is obvious from Andreev reflection~\cite{Petrov07} and shot-noise measurements~\cite{Zhou19}, with evidence of pairs above T$_c$ in the latter case, which we take as a consequence of a short-ranged pairing mechanism. No divergence of the effective mass can be inferred from penetration-depth measurements~\cite{Prozorov06}. Cuprate SC is of the second kind, exhibiting well-defined vortex lattices with BCS-vortex bound states~\cite{Berthod17}. Its somewhat extraordinary properties are a $d$-wave symmetry of the superconducting gap, a coherence length of the order of a nanometer, and consequently an extremely high second critical field. Notably, the Cooper pair size is smaller than the distance between pairs, which is the opposite regime to the elemental BCS one~\cite{Friedel96}. These experimental facts do not preclude the BCS framework, but make its standard phonon-mediated (or any similar long-range) mechanism, with large retardation effects, highly unlikely. It is then a matter of terminology whether to call such SC BCS, BCS-like, or even non-BCS. We opt for BCS, because the fundamental distinction of the BCS scenario is Cooper pairing, while all the other properties can be relegated to various mechanisms and regimes.

In the effort to understand high-T$_c$ SC, it is thus the most plausible, if not the only natural, approach to search for an alternative mechanism for Cooper's scattering instability, identified by BCS. We use the standard term "Cooper pairing" for it throughout, only noting here that it leaves the issue of pair binding ambiguous in the high-T$_c$ context. The Cooper pairs in this work are standard BCS Cooper pairs~\cite{Mahan90}, which are not bound, a crucial point of difference with respect to those pre-formed pair models, bipolaronic in particular~\cite{Alexandrov96}, where pairs have a net negative binding energy. The size of BCS Cooper pairs is just the range of the scattering which maintains the Cooper instability~\cite{Mahan90}. The mean field induced by that instability was recognized through the work of Bogoliubov to be one of only three generically possible: Hartree, Fock, and Bogoliubov. These correspond to the exactly three ways in which one can factor a two-body interaction~\cite{Negele98}. 

Because of the generality of this construction, attempts to explain any observation of SC outside the path traced by BCS, from the Cooper instability above T$_c$ to the Bogoliubov mean field below it, have a very high bar to pass, and none have succeeded so far, not for want of trying. By contrast, attempts to retrace Gor'kov's derivation of the Landau-Ginzburg equations outside the physical regime which he invoked, having lead and mercury in mind, are to the best of our knowledge absent from the literature. However that may be, claims that BCS cannot explain a particular observation in the cuprates must always be measured against the physical regime assumed by the BCS formula employed, otherwise they default to the generally accepted inferences that the Cooper pairing mechanism is not the same, or that the pairs are small. In particular, once the pair size reaches the lower limit set by the size of the unit cell, it is no longer determined by the BCS mean-field equations~\cite{Friedel96}. We describe the microscopic aspects of a possible SC mechanism in the last part of this work.

\subsection{Normal state}

While the SC state seems to be rather conventional, the difficulty in understanding cuprates is due to the complexity of the normal state. It shows unusual evolution of practically all relevant observables in temperature, doping, and the energy and momentum of various probes. Keeping in mind that superconductivity in cuprates is a high-temperature phenomenon, it is natural to approach the normal state properties by identifying the (low-)energy modes which would be excited above 100 K, which also means that even lower energy scales are not of interest.

Attempts to do so all encounter the same obstacle: the cuprate superconductors are primarily charge-transfer ionic insulators, and doping them does not affect the large energy scales responsible for Coulomb binding. It was in fact a motivating conjecture of their discoverer that these Coulomb scales would give rise to a large electron--phonon coupling~\cite{Bednorz87}. The question, how the underlying ionic background influences the metallic carriers, has been central to both the normal state and the SC mechanism of cuprates ever since.

For example, the finding of $T^2$-dependence of the resistivity up to room temperature~\cite{NBarisic13,Ando04a}, discussed below, implies that the electron-electron scattering contributions are not only much larger than k$_\mathrm{B}$T$_{\mathrm{room}}$, but also the lowest of all energy scales affecting transport. Similarly, it is well known that the Cu $3d$ intraorbital (Hubbard) repulsion U$_d$ in cuprate systems is $5$--$10$ eV and that the Cu--O charge-transfer gap $\Delta_{pd}$ is $2$--$3$ eV. Thus, to understand the normal state properties one needs to properly identify and treat the largest energy scales first. Only starting from them can one hope to understand the emergent phases, including high-T$_c$ SC, or properties that appear on even smaller energy scales. For example, the Fermi energy of the pocket in the reconstructed phase at 10\% doping~\cite{Doiron-Leyraud07,NBarisic13a} is of the order of $\sim 10$~ meV~\cite{Doiron-Leyraud15,Tabis21}, which means that it is not a relevant ground state for the SC.

\begin{figure*}[ht!]
\begin{center}
\includegraphics[width=16cm]{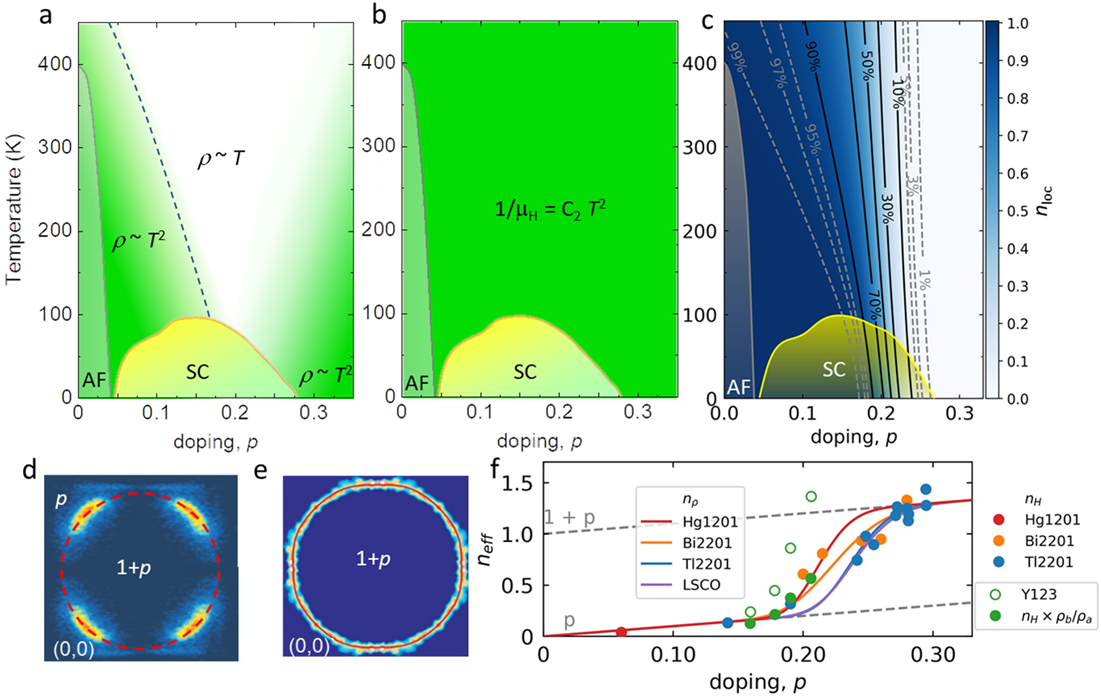}
\end{center}
\caption{(a), (b) $\&$ (c) Phase diagrams summarizing the essence of cuprates. Notably, (a) and (b) are obtained directly from experiment. While (a) the compound independent sheet resistance in cuprates evolves gradually across the phase diagram, from (Fermi-liquid) quadratic to linear-like temperature dependence and back to (Fermi-liquid) quadratic~\cite{NBarisic13}, the Hall mobility (b) is doping- and compound-independent ($1/\mu_H=C_2\,T^2$, where $C_2=0.0175(20)\,K^{-2}$)~\cite{NBarisic19,NBarisic15}. The universality of the Hall mobility, and the unambiguous determination of the Fermi-liquid nature of the itinerant charge (both in the pseudogap and the overdoped regime - green areas in (a)) implies that Hall mobility measures the (doping-independent) temperature dependence of the scattering rate~\cite{NBarisic19,NBarisic15}. Consequently, the gradual evolution of the resistivity is due to the change in carrier density ($n_{\mathrm{eff}}$), which is directly measured by the resistivity (or Hall coefficient)~\cite{NBarisic13,NBarisic19,NBarisic15,Pelc19}. Because the parent compound is a Mott charge-transfer insulator (exactly one hole is localized per CuO$_2$ unit: $n_{\mathrm{loc}} = 1$), the overall charge density is always $1+p = n_{\mathrm{loc}} + n_{\mathrm{eff}}$, where the change in  $n_{\mathrm{loc}}$, and concomitantly in $n_{\mathrm{eff}}$, is due to a gradual delocalization of the Mott-localized hole, with temperature and doping. The obtained variation of the localized hole content is shown in (c)~\cite{Pelc19,Tabis21}. Observe that the $97$\% line coincides with the T$^*$ line in (a). (f) Variation in the measured $n_{\mathrm{eff}}$ (either from  resistivity: $n_{\mathrm{\rho}}$ - full lines~\cite{Pelc19}, or from Hall coefficient: $n_{\mathrm{H}}$ - circles~\cite{Putzke21}), clearly transitioning from $p$ to $1+p$ between maximal and zero T$_c$ in the overdoped region in (c). The apparent discontinuity of the open green circles (Y123)~\cite{Badoux16} disappears when chain anisotropy is taken into account (full green circles)~\cite{Putzke21}. (d-e) The ARPES~\cite{Norman98,Hossain08,Plate05} or STM~\cite{McElroy03} measurements reveal that indeed a partial gap develops on the Fermi surface. The arc lengths observed in (d) are just $p/(1+p)$ of the total length of the Fermi edge, which encompasses the underlying surface $1+p$ [dashed line in (d) and full line in (e)]. Finally, it should be noted that the superconductivity vanishes  by disappearance of either $n_{\mathrm{eff}}$ or $n_{\mathrm{loc}}$~\cite{Pelc19}.
\label{fig2-transport}}
\end{figure*}

The cuprates exhibit a large range of material-depen\-dent phenomena, such that the extraction of those relevant for the SC mechanism is itself a daunting task. There are two main distinctions to be made. One is between hole- and electron-doped materials. The other is between the 100~K T$_c$'s (high high-T$_c$ SC) and the $<50$ K T$_c$'s (low high-T$_c$ SC). Only insights valid across these distinctions have a chance of explaining cuprate SC as a universal phenomenon. We believe that such an explanation is possible, and use the occasion of this Festschrift to describe our reasons for that belief in a uniform and systematic manner. Our scenario is summarized in the captions to two figures, Fig.~\ref{fig2-transport} for the normal state and Fig.~\ref{Figure_5} for SC. We are especially glad that the insights of K.~A.~M\"uller himself, accumulated over several decades after his transformational discovery~\cite{Mueller05,Mueller07}, prove to be relevant and important from our point of view.

\section{Macroscopic features: experimentally established universalities}

Conductance properties correspond to a weighted integration over the whole Fermi surface, probing it delicately within a k$_\mathrm{B}$T energy window. Because of that, they are often last to be understood, although usually first to be measured when mapping the phase diagram. Therefore, in light of the complexity of the cuprates as chemical compounds, it is rather surprising that the sheet resistance (resistance per CuO$_2$ plane) is essentially universal~\cite{NBarisic13}. This observation means that most of the collective compound-specific ordering tendencies observed by various techniques in cuprates do not affect the itinerant charges. Instead, they should be associated with a second subsystem, consisting of a localized charge, as discussed in detail in the following subsections.

\subsection{Fermi liquid nature of the itinerant charge}

At high hole-dopant concentrations $p$, the behavior of cuprates corresponds to that of a conventional Fermi liquid. The Fermi surface is well-defined~\cite{Vignolle08}, the Wiede\-mann-Franz law is obeyed~\cite{Proust02}, resistivity exhibits a quadratic temperature dependence~\cite{Kubo91,Manako92} with a carrier density that corresponds to $1+p$~\cite{Plate05,Hussey03,Proust02}. As the doping decreases, cuprates exhibit an unusual evolution of these properties. For example, close to optimal doping resistivity very gradually evolves from quadratic to linear-like~\cite{Mackenzie96}, down to low temperatures~\cite{Fiory89}. Reducing the doping further, resistivity at high temperatures remains linear-like, while below a characteristic, so called pseudogap temperature T*, the quadratic temperature dependence is reinstalled, as shown in Fig.~\ref{fig2-transport} \cite{NBarisic13,Ando04a}. Notably, in the pseudogap regime the Hall coefficient is independent of temperature and measures the carrier density $p$ throughout it~\cite{NBarisic19,NBarisic15,Ando04}. Both observations strongly suggest that itinerant charges in the pseudogap regime behave as a Fermi-liquid, except that there are $p$ of them. The rather poorly defined concept of a non-Fermi liquid only reflects a confusion over how many carriers there really are. By contrast, the Fermi liquid is well enough defined for its experimental signatures to be precisely identified. The robust demonstration of two textbook Fermi-liquid scalings, namely Kohler's rule~\cite{Chan14} and the temperature/frequency scaling of the optical scattering rate~\cite{Mirzaei13}, unambiguously reveals the Fermi-liquid character of the itinerant-charge subsystem present in the pseudogap regime (Fig.~\ref{Figure_2}).

Despite the unusual and complex evolution of all observables in cuprates, the Hall mobility ($1/\mu_H=\rho/R_H$) behaves astonishingly simply. It remains unchanged across the normal state of cuprates, being essentially independent of regime or cuprate compound in which it is measured, cf.\ Fig.~\ref{fig2-transport}~\cite{NBarisic19,Li16}. This striking experimental fact should be an anchor for any interpretation of the normal state of cuprates. It was established rather early that the Hall mobility is quadratic in temperature at optimal doping ($1/\mu_H=C_2T^2$)~\cite{Chien91}, in contrast with the linear resistivity, but what passed unnoticed was that the coefficient $C_2$ remains the same in electron and hole, underdoped, optimally doped, and overdoped cuprates, thus it is universal~\cite{NBarisic19,Li16}. Notably, the quadratic temperature dependence was also observed in the case of the heavy-fermion system YbRh$_2$Si$_2$ in the proximity of field-tuned quantum criticality~\cite{Paschen04}. Early on, this behavior was taken as an indication that the longitudinal and transverse scattering rates were different, and was associated with spin-charge separation~\cite{Anderson91}. Yet in cuprates, exactly the same Hall mobility is observed from the underdoped to the overdoped limits of the SC dome, clearly indicating that transport itself is not exotic anywhere. In a Fermi liquid, the meaning of the transport coefficients is well-defined [Hall coefficient: $R_H=1/(n_{\mathrm{eff}}e)$; Hall mobility: $1/\mu_H=m^*/(e\tau)$; resistivity: $\rho = R_H/\mu_H$]. With $\mu_H$ fixed and universal, the simplest if not the only possibility is to associate the difference between the under- and overdoped regimes with variations in $n_{\mathrm{eff}}$, which is the basic aggregate variable of pseudogap physics.

\subsection{Charge accounting between two electronic subsystems}

As the term pseudogap~\cite{Friedel88,Alloul89} implies, its primary effect is the reduction of the effective carrier density, while the universality of $\mu_H$ implies that the effective mass and the scattering rate of the charge carriers themselves is unchanged~\cite{NBarisic19}. And indeed, across the cuprate phase diagram, the underlying Fermi surface contains $1+p$ states, as established experimentally by photoemission spectroscopy measurements~\cite{NBarisic19,Pelc19,Drozdov18}. In the highly overdoped regime, the underlying and actual Fermi surface coincide (Fig.~\ref{fig2-transport}). With decreasing doping, a partial gap develops on the antinodal parts of the Fermi surface gradually affecting exactly one hole per CuO$_2$ unit~\cite{NBarisic19,Pelc19}, leading to the formation of Fermi arcs which still contain $p$ states~\cite{Ando04,Padilla05,NBarisic13,NBarisic19}. Because those arc states are of exactly the same character as the Fermi-liquid states observed in the overdoped regime, the normal state of cuprates is properly called "pseudogapped Fermi liquid"~\cite{NBarisic19}. The whole complexity (or "non-Fermi-liquidity") of cuprates stems from the second (localized) electronic subsystem which delocalizes with temperature and doping, becoming part of the itinerant Fermi-liquid electronic subsystem by gradually repopulating the gapped part of the Fermi-surface~\cite{NBarisic19,Pelc19}. Notably, because the Hall mobility ($\mu_H=R_H/\rho$) is independent of doping, the Ioffe-Regel limit in cuprates is never really crossed, but only appears so if one includes the localized hole in the accounting~\cite{NBarisic13}. Thus, there is no need to invoke a quantum-critical regime, in particular because the scattering rate of the truly mobile carriers essentially never changes throughout the phase diagram.

\begin{figure*}
\begin{center}
\includegraphics[width=13cm]{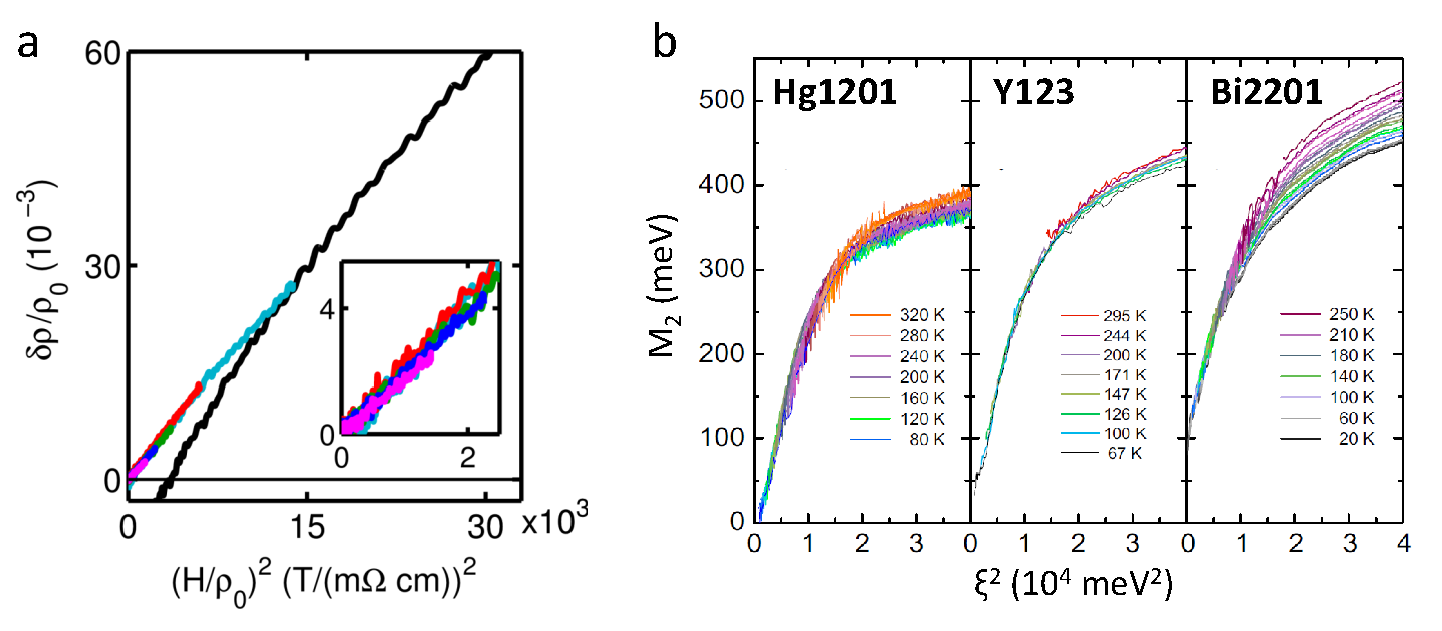}
\end{center}
\caption{
Fermi-liquid scaling of itinerant charges in underdoped cuprates. (a) the Kohler scaling of the magnetoresistivity~\cite{Chan14} and (b) the Fermi-liquid scaling of optical scattering rate $(1/\tau = M_2 \approx (\hbar \omega)^2 + (a \pi k_B T)^2)= \xi^2$, where $a=1.6$, close to its theoretically predicted value 2$)$ ~\cite{Mirzaei13}. Notably, these are the only scalings that have been unambiguously demonstrated in cuprates. Both imply a quadratic temperature dependence of the scattering rate as determined by transport.
\label{Figure_2}}
\end{figure*}

Because the ratio $m^*/\tau$ is fixed by the universality of $\mu_H$, the resistivity $\rho = R_H/\mu_H$ is in fact a direct measure of carrier density $n_{\mathrm{eff}}$~\cite{NBarisic19,NBarisic15}. This fact was recently used in a detailed analysis of the resistivity in different compounds to find the effective carrier density $n_\mathrm{eff}$ as a function of temperature and doping~\cite{Pelc19}. And indeed, as already announced in 2015~\cite{NBarisic19,NBarisic15}, the zero-temperature evolution of $n_\mathrm{eff}$ at $T = 0$ showed a gradual change from $p$ to $1 + p$ (Fig.~\ref{fig2-transport}). Somewhat later, high magnetic field (up to 88T) measurements of Hall number in YBa$_2$Cu$_3$O$_y$ (Y123) yielded a rather sharp change of $n_\mathrm{H}(T = 0)$, from $p$ to $1 + p$, at $p$ = 0.16, which was attributed to a pseudogap critical point~\cite{Badoux16}. The discrepancy was resolved recently by taking chain anisotropy into account, which yielded a smooth crossover for Y123 as well~\cite{Putzke21}. The evolution of the high-field Hall number $n_\mathrm{H}$(T = 0) in two cuprate families, Tl2201 and Bi$_2$Sr$_2$CuO$_{6+\delta}$ (Bi-2201), which have simple single-band Fermi surfaces and do not suffer from the presence of the Lifshitz transition~\cite{Rourke10,Ding19}, is similarly smooth as a function of $p$ throughout the overdoped regime, so that $n_\mathrm{H}(T = 0)$ does not reach the value 1 + $p$ until close to the far edge of the superconducting dome~\cite{Putzke21}, agreeing remarkably well with the evolution of $n_\mathrm{eff}(T = 0)$ determined from the resistivity (Fig.~\ref{fig2-transport}). Once the density of itinerant charges is determined, it is rather straightforward to determine the number of localized holes, because the total charge is always $1+p$:
\begin{equation}
1+p = n_{\mathrm{eff}}+n_{\mathrm{loc}}.
\label{eq-p}
\end{equation}
The established doping dependence of $n_\mathrm{loc}$ is shown in Fig.~\ref{fig2-transport}. We associate the gradual nature of the hole localization process with the well-known evolution of bond-angle disorder in cuprates~\cite{Bianconi87,Egami94,Rullier-Albenque08}. In fact, such a disorder mandates that gaps, associated with the one localized hole per CuO$_2$ unit cell, are also disordered~\cite{NBarisic19,NBarisic15}, because the localization energy should strongly depend on the environment. This disorder results in a 50-meV-broad gap distribution~\cite{Pelc19} and the associated percolation rather than fluctuation processes, which should not be surprising, because Mott localization is by its nature a first-order transition. Moreover, it does not break any symmetry (it is a $\mathbf{q} = 0$ transition) which leaves the Brillouin zone unfolded, so that the underlying Fermi surface does not change and the mobile carriers only experience a crossover in density, manifested by arc growth, instead of any transition induced by the localized subsystem. This dopant-induced arc growth scenario is corroborated by a one-body Fermi liquid DFT calculation~\cite{Lazic15}, as elaborated below.

\begin{figure*}
\begin{center}
\includegraphics[width=17cm]{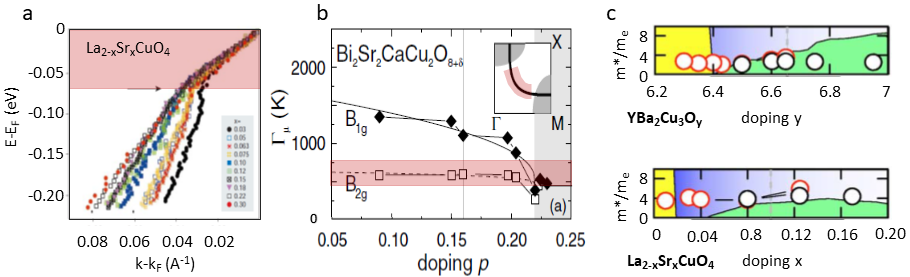}
\end{center}
\caption{Properties of itinerant charges as observed by advanced spectroscopic tools. (a) ARPES reveals the universality (independence of compound and doping) of $v_F$ along the nodal direction up to the energy of 70 meV (indicated by the red area)~\cite{Zhou03}. (b) Doping-independence of nodal ($B_{2g}$) Raman scattering rate ($\Gamma_\mu$) (indicated in red) is also well documented~\cite{Venturini02,Muschler10} (c) Optical spectroscopy not only reveals the Fermi-liquid nature of the itinerant charges (Fig.~\ref{Figure_2}~\cite{Mirzaei13}), but also implies that $n_{\mathrm{eff}} = p$ in the pseudogap regime (not shown~\cite{Padilla05}), and that the effective mass (open circles) is both doping- and compound-independent~\cite{Padilla05}. 
\label{Figure_3}}
\end{figure*}

The above conclusions were made exclusively based on the universalities revealed by transport measurements. Importantly, the results obtained by advanced spectroscopic techniques, partially summarized in Fig.~\ref{Figure_3}, also agree with those conclusions. Regarding the itinerant charges, though spectroscopic techniques are less sensitive than transport in the k${_\mathrm{B}}$T energy window around the Fermi surface, angular resolved photoemission reveals that the nodal Fermi velocity is compound- and doping-independent~\cite{Zhou03}. Optical conductivity indicates both the doping-independence of the effective mass (on top of the Fermi-liquid scaling discussed above and in Fig.~\ref{Figure_2}), and the doping dependence of the density of charge carriers~\cite{Padilla05}, which coincides with the $n_\mathrm{eff}$ obtained from transport~\cite{NBarisic13,Pelc19,NBarisic19}. Finally, Raman spectroscopy reveals the doping-independence of the nodal ($B_{2g}$ symmetry) static relaxation rate~\cite{Venturini02,Muschler10}. 

\begin{figure}
\begin{center}
\includegraphics[width=7cm]{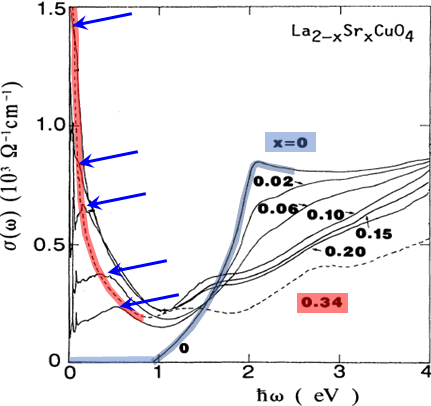}
\end{center}
\caption{Delocalization of exactly one hole per CuO$_2$ unit observed by optical spectroscopy. The parent compound (at x=0) is a charge transfer insulator and a clean gap is observed in optical conductivity (blue-stroked curve). Immediately upon doping, a mid-infrared feature appears. Its maxima for various dopings are indicated with blue arrows. The most important observation is that the mid-infrared feature transfers the spectral weight of the  one localized hole with increasing doping to the coherent (Fermi-liquid) peak at high x=0.34 (red-stroked curve)~\cite{Pelc19,Tabis21,Uchida90}. 
\label{Figure_4}}
\end{figure}

The real advantage of spectroscopic tools in comparison with transport is in capturing the evolution of high-energy scales. And indeed, most if not all of them reveal the presence of the same high-energy scale~\cite{Honma08}. To understand the meaning of this energy scale, it is perhaps most instructive to carefully follow the evolution of the so-called mid-infrared feature in the optical conductivity, shown in Fig.~\ref{Figure_4}. As already pointed out several times in the past~\cite{NBarisic19,NBarisic15,Pelc19}, the energy maximum of this (broad) feature decreases with doping, transferring the spectral weight of exactly one charge per CuO$_2$ unit to the coherent (Fermi-liquid) peak at very high doping. In this way, the same $p$ to $1+p$ change in the carrier density is observed as in transport. Raman spectroscopy also clearly reveals that the "hot" (antinodal) quasiparticles become insulating already deep in the overdoped regime~\cite{Venturini02,Muschler10}, as inferred from the appearance of a dip in the antinodal ($B_{1g}$) part of the spectra. The energy position of the minimum in the dip coincides with the position of the maximum of the mid-infrared feature observed in optics. As already mentioned, ARPES as well as  scanning tunneling microscopy (STM) shows that the (pseudo)gap is present in $k$-space as well as in real space. Once again, positions of the high-energy "hump" in the ARPES spectra coincide with the positions of the mid-infrared feature as a function of doping~\cite{Honma08,Pelc19}. Moreover, the underlying Fermi surface always encompasses $1+p$ states, as corroborated by the well-documented universality of Hall mobility, which excludes a Fermi-surface reconstruction which would fold the zone and thus necessarily change the scattering rate and effective mass~\cite{Tabis21}. The bottom line is that all these spectroscopic tools unambiguously indicate a pseudogap evolving without symmetry-breaking, therefore any discussion of transport in cuprates must take into account the concomitant evolution of the carrier density, although that is regularly ignored.

\begin{figure*}
\begin{center}
\includegraphics[width=12cm]{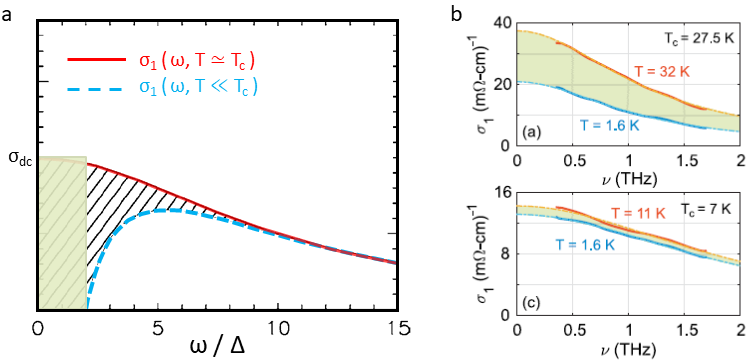}
\end{center}
\caption{
Optical conductivity above and below the T$_c$ in the (a) underdoped (schematically shown~\cite{Homes05}) and (b) very overdoped (as measured~\cite{Mahmood19}) regime. In the underdoped regime it is the itinerant (Fermi-liquid~\cite{NBarisic13,Chan14,Mirzaei13}) charges ($n_{\mathrm{eff}}$) of the normal state that collapse into the superconducting condensate~\cite{Homes04} (note that $n_{\mathrm{loc}}=1$, thus $\rho_S\propto n_{\mathrm{eff}}$). The missing spectral weight which is contained in the SC delta peak at $\omega=0$ is schematically indicated with a hashed area. A rough estimate of the superfluid density, corresponding to the green area, is $\rho_S \propto 2\Delta \sigma_{dc}$, yielding the Homes law (which is usually presented on the log-log graph and thus not sensitive to details)~\cite{Homes04}. The Homes law holds well in the underdoped cuprates and simply states that $\rho_S \propto T_c \sigma_{dc}$ because, according to BCS, $\Delta\propto T_c$~\cite{Homes05}. On the overdoped side, as obvious from (b) it is again the itinerant charges ($n_{\mathrm{eff}}$) which become superconducting, but the measured superfluid density~\cite{Mahmood19,Bozovic16} (corresponding to the green area between the red and blue curve) is proportional to number of localized holes ($n_{\mathrm{loc}}$), which is $\rho_S\propto n_{\mathrm{eff}} n_{\mathrm{loc}}$ according to Eq.~(\ref{eq-rho}) (see Fig.~\ref{Figure_6})~\cite{Pelc19}. Notably, both $n_{\mathrm{loc}}$ and $n_{\mathrm{eff}}$, are obtained directly from the normal state (see Fig.~\ref{fig2-transport}). The reduction of the coherent spectral weight on the overdoped side is only partial simply because there are more itinerant particles than glue (localized holes).
\label{Figure_5}}
\end{figure*}

\subsection{Superconductivity: interplay between the two electronic subsystems}

\begin{figure*}
\begin{center}
\includegraphics[width=14cm]{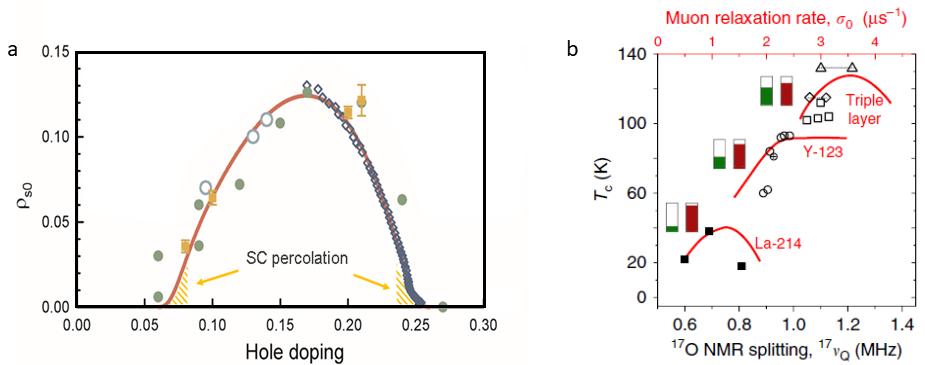}
\end{center}
\caption{
Superfluid density and oxygen-NQR spliting as a function of doping in various compounds. (a) Superfluid density measured in LSCO (open and closed symbols) shows a dome-like doping dependence, which is captured well by Eq.~(\ref{eq-rho}) [full line, corrected at very low superfluid densities (hashed areas) for percolation effects, as discussed in detail in Ref.~\cite{Pelc19}]. (b) The NQR splitting measures the averaged-in-time hole occupancy of the oxygen, while the muon relaxation rate corresponds to the superfluid density. The Uemura plot relates T$_c$ to the muon relaxation rate (solid red lines), revealing its linear-in-$p$ dependence in the underdoped regime~\cite{Uemura89}. Remarkably, similar tendencies as observed in the Uemura plot are obtained by plotting T$_c$ versus planar oxygen quadrupole splitting $^{17}\nu_Q$ (black symbols, lower abscissa). Universality of transport properties implies that the doped $p$ itinerant holes always occupy the oxygen orbitals identically. However, the distribution (static and/or dynamic) of the one localized hole within the CuO$_2$ unit depends on the compound. In the underdoped regime, all compounds show an approximately linear $p$-dependence of electric-field gradients, as well as of the superfluid density (according to the Eq.~(\ref{eq-rho})), associated here with the universal itinerant charges, while the constant material-specific term defines the SC gap $\Delta$ and consequently $\rho_S$. The green (red) boxes denote the electron content of the Cu (O) site, with the white part of the box indicating hole content. Evidently the lower-T$_c$ compounds are closer to Cu $3d^9$.
\label{Figure_6}}
\end{figure*}

We turn to superconductivity now. Direct observation from the optical conductivity in underdoped regime in Fig.~\ref{Figure_5} is that the itinerant carriers become superconducting~\cite{Homes04}. This unambiguously follows from the fact that superconducting state gaps the coherent (Fermi-liquid~\cite{NBarisic13,Chan14,Mirzaei13}) part of the spectra~\cite{Homes04}. On the overdoped side (see Fig.~\ref{fig2-transport}) superconductivity disappears concomitantly with the disappearance of the localized charge. Thus the conjecture that the localized charge is responsible for providing the glue~\cite{Pelc19}. Accordingly, it was proposed~\cite{Pelc19} that the superfluid density, $\rho_S$, is simply proportional to both the density of itinerant holes $n_\mathrm{eff}$ and density of localized holes $n_\mathrm{loc}$:
\begin{equation}
\rho_S = n_{\mathrm{eff}}\:(O_S\:n_{\mathrm{loc}}),
\label{eq-rho}
\end{equation}
where the densities of both electronic subsystems ($n_\mathrm{eff}$ and $n_\mathrm{loc}$) can be directly identified from the normal-state charge transport properties, while $O_S$ is a compound-dependent constant. And indeed, despite this simplicity, a remarkable agreement with experimental data is obtained, shown in Fig.~\ref{Figure_6}.

Notably, such a simple relationship between $\rho_S$ and $n_{\mathrm{loc}}$ implies: (1) that the latter measures the effective SC coupling strength ("glue"), and (2) a very local (and fast/short-time) superconducting mechanism, in agreement with the notion that the size of the Cooper pair (coherence length) in cuprates is of the order of a unit cell. Consequently, superconductivity emerges from the normal state through a percolative process characterized by a (yet again) universal (emergent) energy scale of about $30$--$40$~K~\cite{Pelc18,Yu19a,Popcevic18}. Recently reported THz response measurements in the overdoped regime (see Fig.~\ref{Figure_5})~\cite{Mahmood19} are in full agreement with the picture proposed above. Again, the Fermi-liquid charges condense in the SC state, with the superfluid density (Fig.~\ref{Figure_6}) that corresponds to Eq.~(\ref{eq-rho}). Therefore, in the overdoped regime the doping-dependence of $\rho_S$ is predominantly due to changes in $n_{\mathrm{loc}}$~\cite{Pelc19}.   

\subsection{Degeneracy of oxygen orbitals}

Finally, it should be noted that there is a series of experiments revealing the importance of the degeneracy of oxygen orbitals for the superconducting mechanism. For example, it is well known that the LTT phase in lanthanum-based cuprates is associated with a small structural tilt of the rigid CuO$_6$ octahedra (which include apical oxygens) that keeps the two opposite (O$_x$ $ or $ O$_y$) planar oxygen atoms within the plane while the other two are rotated outside the plane. Remarkably, this small structural change has devastating consequences on the superconducting state, sharply suppressing T$_c$~\cite{Moodenbaugh88,Sera89}. Very early on, it was pointed out that this small LTT essentially does nothing else but lift the pairwise degeneracy of the planar O atoms, changing the intracell Cu--O and O--O charge transfers while keeping the total charge of the CuO$_2$ unit conserved~\cite{Barisic90}. It was also noted that those charge transfers may actually occur by real charge motion (intracell current). Finally, it was concluded that this charge fluctuation is required by the superconducting mechanism~\cite{Barisic90}. Remarkably, it was demonstrated later that the well-known dramatic suppression of superconductivity by Zn impurities in the copper-oxygen plane also occurs by a local breaking of the oxygen symmetry~\cite{Pelc15}. One can understand the effect of the much-discussed charge density wave (CDW) order in the underdoped regime (e.g., in Hg1201 and Y123) which reconstructs the arc-like Fermi surface to a pocket~\cite{Badoux16,Tabis21} on a similar footing. Namely, it is associated with oxygen orbitals~\cite{Comin15}, affecting only 0.03 holes per CuO$_2$ unit, however it is extremely efficient in suppressing superconductivity, presumably again by breaking the symmetry of O$_x$ and O$_y$ orbitals~\cite{Tabis21,Pelc19}. Thus, these three different kinds of experiments all indicate that the oxygen-oxygen charge transfer within the unit cell is important, and that a fluctuating oxygen-copper-oxygen polarization could play an essential role in providing the glue. In this respect, NQR experiments measuring the local electric field gradient, which is highly sensitive to the local charge symmetry, are very instructive. Those measurements show that T$_c$, and concomitantly the superfluid density, are a function of the planar $^{17}$O quadrupole splitting~\cite{Rybicki16}. This was associated with the (time-averaged) hole content of the planar O $2p$ orbital, transferred from the Cu $3d_{x^2-y^2}$ orbital~\cite{Rybicki16}. It is also experimentally established that doped holes are added to the oxygen $p$-orbitals, while the localized hole is found on the $d$ orbital of Cu in the parent compound, and indeed the quadrupole splitting in the underdoped regime shows a universal linear-like in $p$ dependence on top of a material-dependent constant. The latter can be deduced from Fig.~\ref{Figure_6}b by noting that the positions of the muon-relaxation maxima at different values of the splitting occur at similar values of $p$ for the various materials. We attribute the constant part to the intracell Cu--O charge distribution of the localized hole. The higher the splitting of the two oxygens (whether caused by the LTT tilt, bond-angle disorder, symmetry of a particular compound, oxygen CDW, etc.), the lower the SC gap $\Delta$ and T$_c$. The main effect of this splitting is to suppress the O$_x$--O$_y$ charge fluctuation, which thus emerges as a critical ingredient of the SC mechanism. Around optimal doping, hole melting is accompanied by a redistribution/hybridization between the Cu and O orbitals, which is then observed by a change in the doping dependence of the electric field gradient and the superfluid density. It is possible that the distinctions between low- and high-T$_c$ cuprates stem from the fine-tuning of the $p$--$d$--$p$ fluctuation by the Cu-localized holes visiting the neighboring planar-oxygen atoms, which is the reason for the material-dependent constant $O_S$ in Eq.~(\ref{eq-rho}).

\section{Microscopic features}

\subsection{Orbital symmetries and the Coulomb interaction}

The outstanding spectroscopic feature of the cuprates is that a single band crosses the Fermi energy in the planar Brillouin zone (BZ). This apparently simple situation is complicated by the Fermi arcs~\cite{Yoshida12}, which seem to defy understanding within a one-body tight-binding model. The authoritative tight-binding description of cuprates was provided in terms of a six-band model, consisting of two distinct sets of orbitals~\cite{Pavarini01}. Three are the Cu $d_{x^2-y^2}$, O $p_x$, and O $p_y$ orbitals already introduced in the planar three-band Emery model~\cite{Emery87}. The other three are the Cu $4s$, Cu $d_{z^2}$, and apical O $p_z$ orbitals, which present to the plane as a single axially symmetric "effective $s$-orbital," which we call the "$s$-complex" here to allow for more general situations, e.g.\ in the mercury cuprates, where the apical oxygen is separated from the Cu-O plane by a Ba layer, or in the T' phase of NCCO~\cite{Tsukada05,Adachi13} and infinite-layer cuprates, where it is absent.

Because the $s$-complex appears as a single orbital in the plane, an effective four-band model, consisting of the three planar bands and a Cu $4s$ band, has the pleasing property of being simultaneously two-dimensional and containing the orbital responsible for electronic coupling to the perpendicular direction. The most interesting symmetry of this model is that the secular determinant factorizes along the diagonals ($k_x=\pm k_y$) of the BZ in such a  way that the wave function on the conduction-band diagonal does not depend on the $s$-orbital parameters~\cite{Lazic15}. In other words, the BZ diagonal is orthogonal to the $s$-complex in wave-function space, so that the nodal fermions are truly two-dimensional, no matter how strongly the antinodal regions couple to the $s$-orbital.

The latter coupling exposes the antinodal fermions to the Coulomb fields of the interplanar spacer ions. Thus they are part of the 3D charge-transfer insulator, i.e. gapped, by default. This effect alone is able to account for the Fermi arcs within a strictly one-body DFT+U model, as soon as dopants are included as real physical atoms~\cite{Lazic15}. The key observation is that the observed BZ is actually a dopant-disorder-averaged unfolding of a much smaller zone, corresponding to the large unit cell including the dopant ions. This unfolding, originally developed for disordered alloys~\cite{Popescu12}, is trivial in principle but computationally quite demanding. Once carried out, it gives the arcs as observed, including arc growth with doping, without the need for any intraorbital  correlations. Thus the carriers are easily shown to be a Fermi liquid despite the appearance of the arcs, which are due to both out-of-plane ion fields and the large Hubbard U$_d$ on the same footing. The key role of the latter is in the localization of the hole (the $1$ in $1+p$), which determines the size of the gap (and the binding energy~\cite{Hautier12}). With respect to these effects, correlations of the nodal carriers (the $p$ in $1+p$) are negligible to first order~\cite{Lazic15}. Notably, a similar separation of roles of itinerant and localized carriers was studied in the context of SmB$_6$ and transition-metal oxides already by Falicov and Kimball~\cite{Falicov69}.

The main quantitative effect of the $s$-complex on the planar electrons is that it facilitates the intracell O $p_x$--$p_y$ hopping via a second-order virtual process $2p_x$--$4s$--$2p_y$, which can be estimated as $t_{ps}^2/\Delta_{ps}\sim 0.7$--$1$ eV, nearly an order of magnitude larger than the direct chemical O$_x$--O$_y$ overlap $t_{pp}\sim$ 0.1 eV. Importantly, because the $s$-complex is isotropic in the plane, this hopping has the same symmetry as the direct $p_x$--$p_y$ hopping, so the respective amplitudes add coherently.

\subsection{A doping mechanism}

As is well known, the triply-ionized Cu$^{3+}$ ion ($3d^8$ configuration) is energetically very unfavorable, which is taken into account in the tight-binding formulation as a strong repulsion (Hubbard U$_d$) prohibiting two holes in the Cu $d_{x^2-y^2}$ orbital. By contrast, the Cu$^{+}$ ($3d^{10}$) configuration is easily accessible and provides a plausible doping mechanism in the cuprates~\cite{Mazumdar89}. Dopant ions in the spacer layers provoke a "Coulomb domino effect" in the dielectric layer by which the planar Cu$^{2+}$ ion of the parent insulating compound is reduced via an orbital transition:
\begin{equation}
\mathrm{Cu}^{2+} + \mathrm{O}^{2-} \rightleftharpoons \mathrm{Cu}^{+} + \mathrm{O}^{-}.
\end{equation}
The ensuing hole on the oxygen then delocalizes via O$_x$--O$_y$ hopping, providing the carriers which metallize the planes, eventually superconducting.

Two key predictions of this mechanism have been confirmed, one by direct measurement~\cite{Pelc15}, which shows that substitution of Cu by Zn reverses the orbital transition locally, and the other by calculations~\cite{Lee06,Lazic15}, which show that the dopant charge remains localized in its vicinity, i.e.\ does not reach the planes physically---the charge redistribution around the dopant is found to be neutral overall within a single spacer layer. At first sight, the above metallization of the Cu--O ionic bond would give rise to a mixture of Cu$^{2+}$ and Cu$^{+}$ configurations in the plane, bathed by a Fermi liquid of oxygen hole carriers. However, such a mixture cannot maintain the sharp Coulomb discontinuity between the Cu$^{2+}$ and Cu$^{+}$ states. In the original proposal, this issue was resolved by claiming that \emph{all} Cu $d$-orbitals close~\cite{Mazumdar89}. Based on later direct observations, as discussed thoroughly in the previous sections, we prefer the more moderate picture, where the mismatch is relaxed through the Cu--O overlap into an intermediate uniform Mott-like-covalent-metallic state, which evolves with doping, but retains a significant part of the localized holes on the Cu $3d$ orbital. The doping evolution affects the lattice as well, through the Cu-apical O bond distance and plane buckling (bond disorder). The mechanism of high-T$_c$ SC necessarily plays out in these material surroundings.

\subsection{The $s$--$p$ and $d$--$p$ transport channels}

The O hole in the doped Cu--O plane has three ways to propagate. One is direct O$_x$-O$_y$ hopping $t_{pp}\sim 0.1$ eV, which is the smallest of the three overlaps involved, so we neglect it to first order. The other is the $t_{pd}$ overlap with the Cu 3$d_{x^2-y^2}$ orbital, with hopping suppressed by the Hubbard repulsion when it is in the $d^9$ configuration, and by the Pauli principle when it is in the closed $d^{10}$ configuration. The only unhindered path is the previously-mentioned effective $t_{pp}\sim t_{ps}^2/\Delta_{ps}\sim 0.7$--$1$ eV. The need for such a large effective $t_{pp}$ to fit ARPES in the three-band model (without the $4s$ orbital) may be taken as direct proof that the $s$--$p$ channel is active at the Fermi level, even in the electron-doped compounds~\cite{Sunko07}. It was pointed out~\cite{Pavarini01,Andersen01} that the effective $s$-orbital needs to be "nearly pure Cu $4s$," i.e. unhybridized with other orbitals in the $s$-complex, for T$_c$ to be large.

The localized hole observed experimentally does not all reside on the Cu 3$d_{x^2-y^2}$ orbital (see Fig.~\ref{Figure_6}b)~\cite{Rybicki16}. The pure $3d^9$ (Cu$^{2+}$) configuration is characteristic only of the parent compound. The phenomenology is simply that there is a Coulomb blockade preventing the itinerant O $2p$ hole from hopping onto the Cu $3d$ orbital, so it should propagate primarily through the $4s$ orbital complex. The localized hole is dynamically shared between the Cu and O sites, and only the way in which it blocks hopping of the itinerant hole onto the $d$ orbital evolves with doping. At the extreme underdoped side, with Cu close to $3d^9$, the reason is the on-site Hubbard repulsion. Near optimal doping, the localized hole is still fully localized within the unit cell, but visits the oxygen orbitals as well, which still blocks, now kinematically, the itinerant hole from visiting the $d$ orbital. It is localized because the Cu--O $d$--$p$ bond is ionic all the time, so the static charge on the $3d$ needs a static countercharge on the $2p$. The total charge $1+p$ then means that exactly $p$ O $2p$ holes per unit cell are free to act as a Fermi liquid via the coherent $2p$--$4s$--$2p$ hopping.

The localized hole (one per CuO$_2$ unit cell) persists well beyond optimal doping, its concentration tapering off gradually (Fig.~\ref{fig2-transport}). To understand the overdoped side in the above scenario, one must distinguish ionic and covalent bonding. Neither is intrinsically metallic. In ionic bonding, an electron is transferred from one atom to another, where it stays. In covalent bonding, a pair of electrons is shared between neighboring atoms. They are still localized to zeroth order in the space of the bond.

In cuprates, the ionic localization mechanism is more involved than in the description above, which literally pertains to rock salt. As is well known, the half-filled band is insulating because the fluctuation of the Cu $d^9$ configuration, required for conductivity, is highly asymmetric, with the $d^8$ (Cu$^{3+}$) suppressed by the Hubbard U$_d\sim$~$5$--$10$~eV. Physically, one can imagine the hole trying to escape in a high-frequency oscillation, but never getting past the first O atom. Upon doping, this local "molecule" is observed as a mid-infrared feature in optical conductivity (Fig.~\ref{Figure_4}). This scenario is usually called Mott localization, although Mott himself~\cite{Mott68} gave credit elsewhere~\cite{Boer37}. It has two variants, charge-transfer when $\Delta_{pd}\ll U_d$ (the cuprate case), and Mott-Hubbard when $\Delta_{pd}\gg U_d$ (not relevant here). Upon doping, an off-site Coulombic mechanism is activated~\cite{Mazumdar89}, in which the Madelung energy gain from the dopants pushes the hole from the Cu onto the O to a varying degree, as described above. We call the whole Mott-charge-transfer-Madelung scenario "Mott-like." The salient point is that as long as the hole is localized at the level of a unit cell, it is not available for band conduction, no matter how it is shared among the constituents.

Both Mott-like (e.g., in cuprates) and covalent bonds (e.g., in polymers) can metallize upon doping as a first-order effect, without affecting the condensation energy to zeroth order. The distinction we make between the two cases in doped cuprates is that the Mott-like (ionic) bond metallizes by diffusion, so conduction is incoherent. The metallized covalent bond is coherent and essentially the same as in elemental metals. (In the latter, the itinerant carrier is also responsible for the condensation energy.) Thus the $d$--$p$ transport channel is diffusive, while the $s$--$p$ channel is dispersive.

Because the actual lattice ions do not diffuse as such, diffusion in the $d$--$p$ channel corresponds to diffusion of the local structures created by the localized hole. It is related to incoherent hopping of the localized carriers. As such, it has been noticed by a number of authors, most notably by K.~A.~M\"uller himself, as described below. It may be interpreted as the movement of small polarons. Generically, we call the object which diffuses "the $d$-complex." It can naturally exceed the Ioffe-Regel limit in transport, because there is no limit in principle on the time a local deformation can reside at any one site in the lattice before moving on. The $d$-complex is diffusive as long as the hole is localized, which are just two ways of saying that the Mott-like Cu--O bond does not participate in coherent (band) conduction. The local structure around one hole need not be compatible with the one around another, creating a large amount of small patches, or short-range HTT/LTT/LTO disorder, which can rearrange themselves with external fields and temperature changes. It has also been proposed that they order as a Wigner crystal, which contains pre-formed zero-coupled pairs~\cite{Mazumdar18}. Others have identified pre-formed pairs with the corresponding bipolarons~\cite{Alexandrov96}. In our view the $d$-complex ($n_{\mathrm{loc}}$) plays an important role in the SC, but is not superconducting by itself, so its diffusion is theoretically a side issue, and experimentally negligible (see Fig.~\ref{fig2-transport} and the related discussion). Therefore, we take the diffusive $d$-complex to be static from now on, removing it from transport accounting in accord with experiment.

\begin{figure}
\begin{center}
\begin{tabular}{cc}
\includegraphics[width=3.5cm]{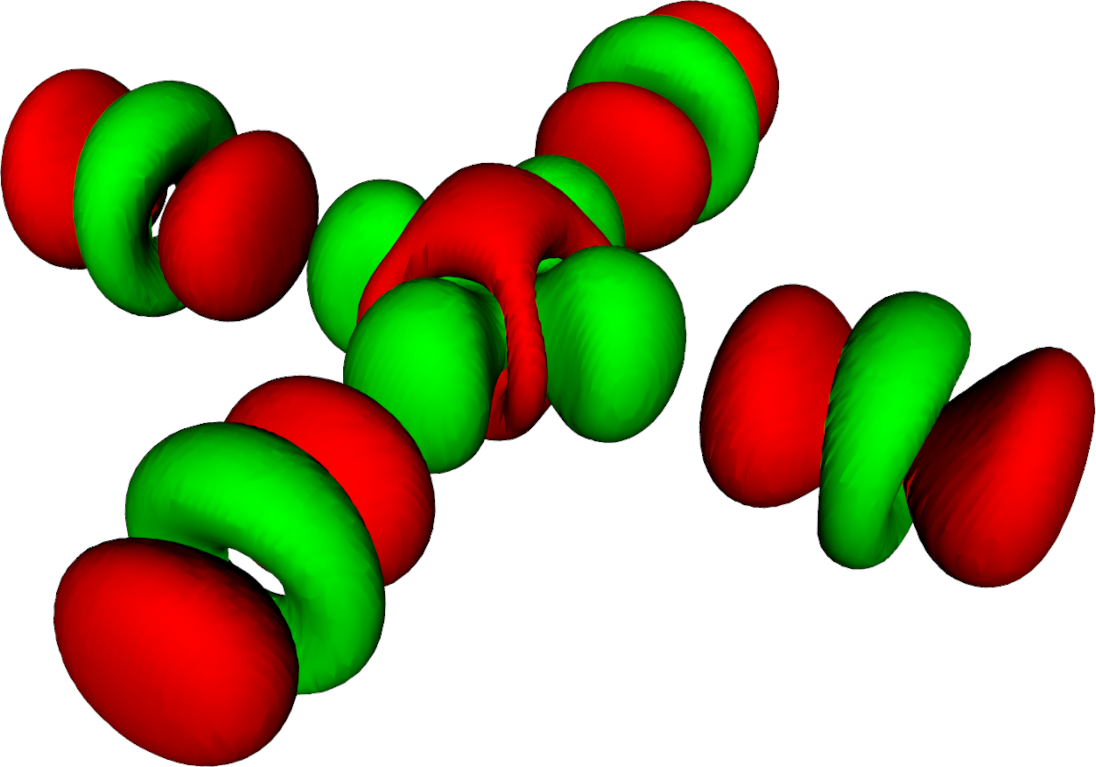} &
\includegraphics[width=3.5cm]{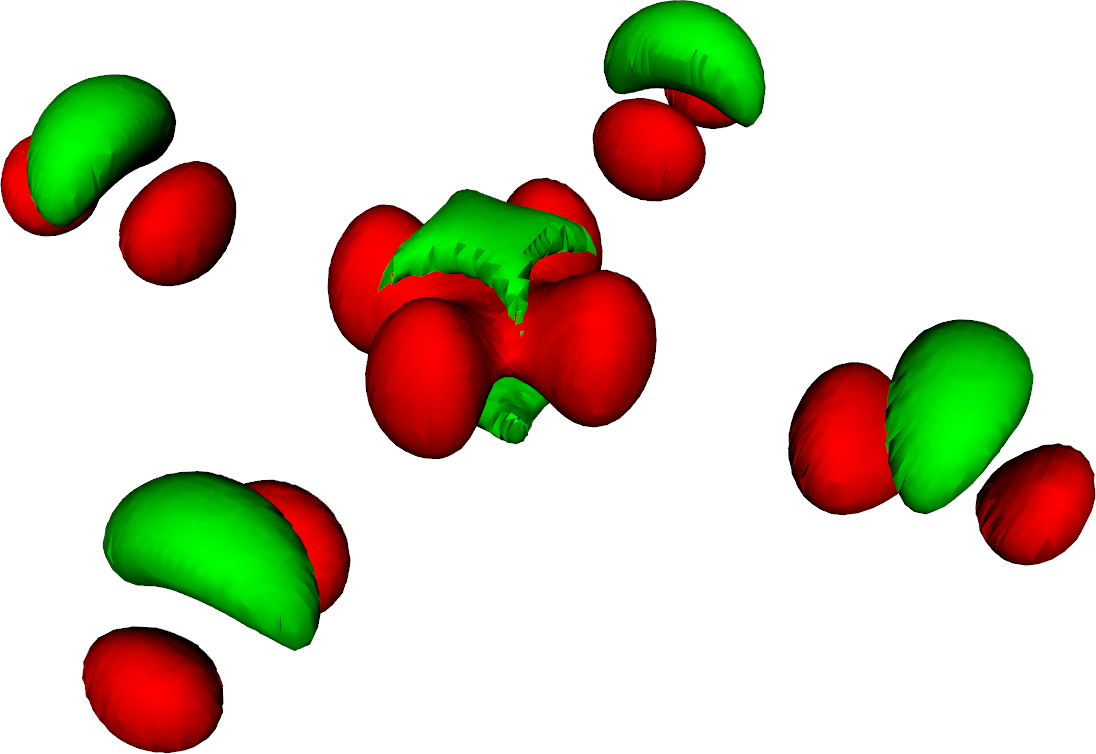}\\
(a) & (b)
\end{tabular}
\end{center}
\caption{(a) Hole doping by physical out-of-plane dopants. The Cu $d$ orbital (green) gives up positive charge (red) to O p. (b) Hole doping by uniform background. Both Cu $d$ and O $p$ orbitals acquire positive charge. Green/red: negative/positive charge difference relative to undoped. After Ref.~\cite{Lazic15}.
\label{fig-orb}}
\end{figure}

On the overdoped side, the Coulomb blockade is relaxed by the large number of carriers in the plane, simultaneously with the Coulomb field of the dopants becoming uniform at the scale of a unit cell. The arcs projected from the dopant zone then touch the edge of the standard zone, as indeed observed in ARPES. The local effect is that the $d$--$p$ bond changes from ionic to covalent, or, in standard language, the $d$ and $p$ orbitals hybridize. The effects of the transition from Mott-like to covalent bonding on the orbital occupation are shown in Fig.~\ref{fig-orb}~\cite{Lazic15}. In panel (a) we show the Cu--O charge transfer occurring when a physical dopant ion is added to the spacer layer: a net positive charge is transferred from Cu to O. The green-red gradient in the figure indicates that this doped charge is Coulomb-blocked from returning to the Cu $d$ orbital. In panel (b), the same doping is realized through a simple shift of the chemical potential, compensated with a uniform charge background. Now the doped charge is shared without blocking: the extra hole is moving onto both the Cu $d$ and O $p$ orbitals. This is the standard situation of a metallic covalent bond, efficiently modeled by tight-binding Hamiltonians.

Case (a) in the figure pertains to the underdoped situation, where the Cu--O bond is Mott-like. The O holes can hop coherently only via the $4s$ orbital, because the $3d$ orbital is blocked. Case (b) is the overdoped situation, where the bond is covalent, and O holes are shared with the Cu $d$ orbital as band fermions. In this limit, the $d$--$p$ channel becomes coherently metallic, i.e. an ordinary Fermi liquid. However, this Fermi liquid does not become superconducting, as we explain now.

\section{A unified narrative}

The picture of high-T$_c$ SC in the cuprates which emerges from the considerations above is that a Fermi liquid of O $2p$ holes propagating via the Cu $4s$ orbital interacts by scattering on the localized $d$--$p$ hole, which plays the role of a "glue." This scattering provides the Cooper-pairing mechanism of high-T$_c$ SC, whose precise microscopic nature remains to be established in full. The evidence collected so far provides important constraints and a number of consequences of that picture, which we summarize here.

The large value of the effective hopping $t_{pp}$ needed to fit ARPES is a strong indication that the $s$-complex is the dominant mode of propagation of O $2p$ holes. The observation~\cite{Pavarini01} that the effective $s$ orbital must be pure Cu $4s$ for T$_c$ to be large implies that the Cooper pairing itself occurs via the same $4s$ orbital. Therefore, the Fermi liquid in the $s$--$p$ channel becomes superconducting. The relaxation of Coulomb blocking by which the $d$--$p$ channel becomes coherent on the overdoped side results in the loss of the localized hole, thus removing the SC pairing mechanism. Consequently, the $d$--$p$ Fermi-liquid channel never becomes SC, although it can carry a large part of the normal transport on the overdoped side. 

The first consequence of this scenario is that the Cooper pairs are very small, of the order of a single unit cell, by default smaller than the distance between pairs, which is the opposite regime to the usual BCS one. Next, because the diagonal of the zone is orthogonal to the $s$-complex in wave-function space, the nodal carriers cannot participate in Cooper pairing, so the SC order parameter is zero on the zone diagonal by symmetry. Third, the same kinematics implies that the non-SC carrier on the zone diagonal will be a nearly perfect low-temperature conductor in the underdoped regime, because hopping via the small direct $t_{pp}$ overlap does not couple either to the out-of-plane dopant disorder, or to the in-plane $d$--$p$ disorder. Fourth, the Cooper pairs can tunnel via the $s$-complex into the neighboring plane, providing a natural path for 2D SC to become 3D. Because the $s$-complex is axially symmetric, such localized tunneling naturally allows for an $s$-wave admixture to the order parameter of 3D cuprate SC, as distinct from its 2D manifestations. Notably, it was pointed out by M\"uller and Keller~\cite{Mueller97} that pure $d$-wave cuprate SC is found only by surface probes, while bulk probes indicate a small $s$ component. Finally, the very large scale $\Delta_{ps}\sim $~$4$--$16$ eV involved in the O--O hopping via the $4s$ orbital means that this virtual process is fast and does not couple to the lattice or any other low-energy perturbation.

The last point naturally explains the large temperature range of the pure $T^2$ scattering, in contrast to the Grüneisen crossover behavior usually observed in metallic conduction~\cite{Garland68,Garland69}, as well as the negligible value of residual resistance~\cite{NBarisic13,Rullier-Albenque08,Ando04a}, essential for observation of quantum oscillations~\cite{Doiron-Leyraud07,NBarisic13a}, in otherwise disordered compounds. Moreover, it also explains the universality of cuprate conduction and SC, because they only depend on the Cu--O bond in the $s$--$p$ and $d$--$p$ channels, which are both universal features of cuprate chemistry. Essentially, the material-\-dependence is in the spacer planes, and stops at the Cu site, manifesting itself on SC only in the amount of sharing of the localized hole between Cu and O. Like the formerly noticed degree of purity of the Cu $4s$ orbital~\cite{Pavarini01}, the material-dependent sharing of the localized hole~\cite{Rybicki16} affects only the value of T$_c$, an important consideration but not a qualitative one.

\section{Discussion}

\subsection{The phenomenology}

The experimentally established and, for cuprates, universal evolution of the transport properties across the phase diagram of hole-doped cuprates (though also relevant for electron-doped cuprates~\cite{Li16,Li19}), discussed above, reveals two electronic subsystems. One corresponds to a (pseudogapped) Fermi-liquid (as demonstrated by two characteristic scaling signatures~\cite{Mirzaei13,Chan14}), while the other is Mott-like localized, or, within the above scenario, it places the local CuO$_2$ bond in the ionic limit.

In large part, the strangeness of these materials lies in the gradual, as opposed to critical, delocalization of exactly one hole per CuO$_2$ unit, driving an underlying inhomogeneity manifested through gap disorder~\cite{Pelc19}. The accompanying change of the effective carrier density, as a function of doping and temperature, from $p$ to  1 + $p$ is apparent from measurements of dc-resistivity~\cite{NBarisic13,NBarisic19,Pelc19} and Hall coefficient~\cite{NBarisic19,Ando04,Putzke21}. The continuing presence of some localized carriers deep in the overdoped regime is obvious not only from transport but also from  scanning tunnelling microscopy~\cite{Gomes07} or Raman spectroscopy~\cite{Venturini02,Muschler10}. It seems plausible that the concomitant change in the density of localized charges [according to Eq.~(\ref{eq-p}) $n_\mathrm{loc}$ changes from 1 to 0, see also Fig.~\ref{fig2-transport}], should be accompanied by considerable configurational entropy in the crossover regime, because cells with localized and non-localized charge (blocked and unblocked $d$-orbitals) are present at the same time. This contribution from the localized subsystem can thus be the reason why the entropy increase as a function of doping~\cite{Loram93,Loram01,Putzke21} is considerably faster than the crossover ($p$ to  $1+p$) seen by resistivity~\cite{NBarisic13,NBarisic19,Pelc19} and Hall coefficient~\cite{NBarisic19,Ando04,Putzke21}. Such effects have been observed both numerically~\cite{ElShawish03} and analytically~\cite{Sunko05} in models based on the Falicov-Kimball scheme~\cite{Falicov69}.

A remarkable feature of cuprates is that the Hall mobility is essentially universal, exhibiting T$^2$ temperature dependence in agreement with the notion that the itinerant charges do not change their Fermi-liquid character, which is clearly identified both in the highly overdoped and pseudogapped regimes. Consequently, the effective mass~\cite{Padilla05}, Fermi velocity~\cite{Zhou03} as well as the umklapp~\cite{Jacko09,Hussey05} scattering rate are doping- and compound-independent~\cite{NBarisic15,NBarisic19,Tabis21}. The whole phenomenology of cuprates can then be encompassed by only two rather simple equations [Eqs.~(\ref{eq-p}) and~(\ref{eq-rho})], where the itinerant charge density is directly read from the resistivity (or Hall coefficient)~\cite{NBarisic15,NBarisic19,Pelc19}. Accordingly, the pseudogap formation is a gradual/percolative process that is nearly complete at the characteristic pseudogap temperature, which leaves nodal $p$ states (arcs) unchanged. The formation of antinodal gaps is thus associated with Mott-like (\textbf{q}=0) localization, which is the highest energy scale in the cuprate problem, contributing to the material binding~\cite{Hautier12}. All the other, often material-specific, orders or ordering tendencies, including superconductivity, should be seen as emergent phenomena. For example, the broken (\textbf{q} = 0) symmetry states observed within the pseudogap, such as $\mathbf{q} = 0$ magnetic order~\cite{Fauque06,Li08}, or electronic nematicity~\cite{Daou10,Murayama19}, might be seen as emergent phenomena related to the charge localization. As already pointed out, the fluctuation of the localized charge might be seen as real charge motion, that is, as an intracell current~\cite{Barisic90}. Acquiring coherency among the cell currents would lead to a current-loop ordered state~\cite{Varma97}, measurable by neutrons~\cite{Fauque06,Li08}.

It is also well established that even a tiny doping causes buckling of the CuO$_2$ planes, which might be seen also as planar bond-angle disorder, or in other words, creation of HTT/LTT/LTO disorder. In our view this is a simple consequence of hole localization which must also be compatible with the itinerant subsystem, which naturally causes a local distortion. Notably, the value of the charge transfer gap is well defined in the parent compound, while the doped material shows a broad gap distribution (Fig.~\ref{Figure_4}). Depending on the overall crystal structure as a result of negotiation/competition of the three involved subsystems (two electronic and lattice) either short-range HTT/LTT/LTO disorder will be present (e.g., Tl2201~\cite{Pelc21}, or nematicity will appear (e.g., Y123~\cite{Hinkov08} or BSCCO~\cite{Lawler10}) or in some cases stripes (e.g., LSCO~\cite{Tranquada95}) which correspond to two different conformations of the CuO$_6$ octahedra assigned to two types of stripes with different (LTO-like or HTT-like) lattices~\cite{Bianconi96}. Finally, it should be mentioned that the currently much discussed CDW correlations seem not to be related to the localized charge, but that they should be primarily associated with the itinerant subsystem. The CDW wave-vector connects the ends of the arcs~\cite{Tabis17} and accordingly reconstructs the Fermi surface, localizing about 3$\%$ of itinerant charges in a CDW ordered state at high fields (above 30 Tesla) and low temperatures (below 20 K)~\cite{Tabis21}. 

\subsection{Insights of K.~A.~M\"uller}

After discovering SC in cuprates, K.~A.~M\"uller had a number of later insights, collected in two reviews~\cite{Mueller05,Mueller07}, which are significant in the frame of our narrative. Most of them are related to the role of the localized hole in our picture. First is the observation of a local deformed structure centered on the Cu site, which appears as a small polaron. We call it the localized hole, or the $d$-complex to emphasize the orbital context. Second is the nature of the local symmetry breaking around it, which can include LTT tilts which suppress T$_c$, moving the material towards a phase-separation instability originally identified as stripes. We would generalize this observation to all kinds of nematic and HTT/LTT/LTO disorder due to the interaction between the localized charge and the lattice. Third is drawing attention to a gigantic O$^{18}$ isotope effect in T*, of the order of 50\%, easily the largest observed anywhere. In our view it should be associated with the localization energy of the hole which deforms the $d$-complex, clearly depending on the electron-phonon coupling with its accompanying mass dependence. The size of the effect can then be simply understood as a consequence of the Coulomb scales involved being very large relative to $T^*$. However, in our scenario the emergent scale associated with SC itself primarily depends on the interplay between the localized and itinerant charges.

The relationship between short-range disorder and the SC T$_c$ has been noticed by others as well~\cite{Phillips03}. In fact we attribute a number of emergent orders observed in cuprates to a localized carrier and its interplay with the structure and/or itinerant charge. In this sense, the insights of K.~A.~M\"uller and the discussions related to polarons~\cite{Mihailovic02,Shengelaya14} are broadly significant.

In addition to the polaronic sector, K.~A.~M\"uller emphasized the existence of a second itinerant-fermion sector~\cite{Mueller98}, which we identify as a Fermi liquid, and which becomes SC in our approach (the $s$-complex). In contrast, he ascribed the SC itself to the polaronic one (our $d$-complex)~\cite{Mihailovic97}. Remarkably from our point of view, he noticed~\cite{Mihailovic97} a rarely cited set of optical measurements~\cite{Holcomb94,Holcomb96} which identified a high-energy ($1.5$--$2$~eV) mode related to SC, which is a natural candidate for the SC glue in our scenario, as explained below. The indication of a small $s$ component in the predominantly $d$-wave order parameter did not escape his attention either, as already noted above~\cite{Mueller97}. Finally, we would like to point out a very early work~\cite{Deutscher87} which argued for the "glassy" nature of SC in cuprates based on the very short coherence length, which we call percolative today.

\subsection{The mechanism}

From the above analysis of transport data across the phase diagram of cuprates, which is also in agreement with the results obtained by advanced spectroscopic tools, it follows that the itinerant Fermi-liquid charge becomes superconducting (by Cooper pairing) while the localized $d$--$p$ hole provides the glue. Such a mechanism is highly local, both fast and short-range. More specifically, several experiments imply that the mechanism should be associated with a charge fluctuation involving both copper and oxygen planar orbitals~\cite{Barisic90,Pelc19,Pelc15,Tabis21}. Accordingly, superconducting properties will strongly depend on the capability of the charge, primarily localized on Cu, to visit the neighbouring oxygens. NMR provides a direct measure of the (total) oxygen occupancy~\cite{Rybicki16} and indeed, the higher the (local) hole occupancy of oxygen, the higher is the superfluid density (and T$_c$). Finally, measurements of the nonlinear conductivity and a novel analysis of the dc-resistivity, Seebeck coefficient, specific heat coefficient, tomographic density of states, and torque magnetometry measurements revealed that SC indeed emerges via a percolative process, universal across the cuprates~\cite{Pelc18,Popcevic18,Yu19a}, in agreement with the very local nature of the glue and the resulting small size of the Cooper pair. Therefore, it is safe to say that possibilities for the SC mechanism in cuprates have been narrowed down considerably after several decades of experimental and theoretical efforts, outlined in this work.

The main issue left open is  the exact microscopy of the local mechanism responsible for Cooper pairing. Since the possibilities are sufficiently limited, the problem of identifying this mechanism is no longer one of a lack of specific ideas. There are only so many Cu orbitals to which the O $2p$ hole can couple. Rather, as already noted in various related contexts~\cite{Anderson07,Mazumdar18}, it is that the large Coulomb scales involved put even a room-temperature SC T$_c$ in the range of a rounding error. In the absence of sufficiently trustworthy computational approaches, one must rely on qualitative arguments to filter particular mechanisms by observation. 

Our starting point is the concept of "glue"~\cite{Anderson07}. Specifically, is the interaction retarded? If it is instantaneous, it is purely kinematic, with the Cu ion a perfectly rigid scatterer. This is the glueless limit. A well-known example of this sort is the superexchange interaction~\cite{Eskes93} $J\sim t_{pd}^4/\Delta_{pd}^3$ which remains when the Hubbard U$_d$ is infinite (perfect rigidity). However, as argued in detail elsewhere~\cite{Sunko20a}, interactions in the spin channel may be relevant for SC in the pnictides (where two electronic subsystems, Fermi-liquid and incoherent, have also been identified ~\cite{Wu10,NBarisic10}), but not for the cuprates. In particular, AF correlations near optimal doping are in the opposite regime in hole- and electron-doped cuprates ($kT_c<\hbar\omega_{\mathrm{magnon}}$ vs. $kT_c>\hbar\omega_{\mathrm{magnon}}$), in contradiction with the idea that the mechanism should be the same~\cite{Sunko07,Li19}. Along the same lines, SC extends all the way to stoichiometry in electron-doped cuprates when AF disappears by outgassing the apical oxygen~\cite{Adachi13,Tsukada05}. We believe that the interaction relevant for SC is in the charge, not spin, channel, so not glueless, but not much retarded either, otherwise the Cooper pair would not be small, or the second critical field large.

If the Cu ion is not rigid, the question arises of its intermediate states which are excited in the Cooper pairing, and which of them provide an attractive interaction. Because Cooper pairs are not bound, theoretically even a repulsive interaction can trigger the Cooper instability~\cite{Maiti13}, an unlikely scenario but not impossible if its range is shorter than the distance between pairs. The relatively conventional nature of the SC state mitigates against it, however. Here, we briefly list reported candidates which are compatible with our ideas in principle. Proposals which require the $d$-complex itself to be superconducting, or act as a reservoir of pre-formed pairs, are incompatible, with a terminological \emph{caveat}. We use the term pre-formed pair to mean a pair with a real negative binding energy. Observations of pairs above T$_c$~\cite{Zhou19} could be islands of small Cooper pairs in our sense~\cite{Pelc18,Popcevic18,Yu19a}, even if others prefer to call them pre-formed. Our comments in the following are meant to sensitize the reader to the issues involved in integrating the physics and chemistry of functional materials, not as an evaluation of the works cited.

Very recently, while considering the evolution of transport coefficients only in the overdoped regime, it was argued that the reduction of $n_\mathrm{H}$(T = 0) and concomitant emergence of linear resistivity in both Tl2201 and Bi-2201, are due to a growth of quasiparticle decoherence on some parts of the Fermi surface~\cite{Putzke21}. Additional arguments along those lines were put forward based on the evolution of high-field magnetoresistance, which suggested that despite having a single band, the cuprate overdoped metal phase hosts two charge sectors, one containing coherent quasiparticles (here corresponding to $n_{\mathrm{eff}}$), and the other ($n_{\mathrm{incoherent}}$) scale-invariant "Planckian" dissipators~\cite{Ayres21,Grissonnanche21}. In the latter, quasiparticle decoherence is introduced as the linear dependence of the decay rate on temperature, energy and an associated timescale conjectured as the shortest time in which energy can be dissipated~\cite{Hartnoll15,Zaanen19}. The conclusions~\cite{Ayres21}, regarding the charge density and superfluid density, are very similar to those summarized by Eqs.~(\ref{eq-p}) and ~(\ref{eq-rho}), proposed earlier~\cite{Pelc19}. The only difference is that in both equations $n_{\mathrm{loc}}$ is replaced by $n_{\mathrm{incoherent}}$, while in Eq.~(\ref{eq-rho}), changes in $n_{\mathrm{eff}}$ are disregarded, which is a good approximation only in the highly overdoped regime, where $n_{\mathrm{loc}}$ is nearly zero (see Fig.~\ref{fig2-transport}). Consequently it was concluded that the incoherent, not the coherent (Fermi-liquid), carriers are involved in the superconductivity, while the underdoped regime was not discussed. However, it is precisely in the underdoped regime that this issue is clearly resolved. In particular, it is obvious there that coherent (Fermi-liquid ~\cite{NBarisic13,Chan14,Mirzaei13,NBarisic19,NBarisic15}) carriers are the only ones present (Figs.~\ref{fig2-transport}(a) and ~\ref{Figure_2}), and that they become superconducting ~\cite{Homes05} (Fig.~\ref{Figure_5}).

Among the candidate intermediate states, the only ones which were claimed to be identified by direct observation involve the Cu $t_{2g}$ $d$-orbitals~\cite{Holcomb94,Holcomb96,Little07}, usually neglected in this context, but noticed by K.~A.~M\"uller nevertheless, as described above. Indeed, one cannot exclude the two local Cu $d_{x^2-y^2}$--$t_{2g}$ excitons of $1.2$ and $1.7$~eV, reported there~\cite{Little07} (or at somewhat different energies reported later, e.g.,~\cite{Chaix17}), as a possible glue in cuprate SC. In fact, our scenario requires that the glue involve a high-energy local process, which is turned off when the Cu~$3d_{x^2-y^2}$--O~$2p$ bond becomes covalent, so alternative Cu $d$-orbitals are a reasonable possibility. Notably, the observed energy scales of the $t_{2g}$ excitons are of the same order as the mid-infrared feature connected with the localized hole, so they could be potentially associated with that hole.

All other candidates of which we are aware are theoretical. None of them take into account the crucial role of the localized hole for the SC mechanism, or of the transition from Mott-like ionicity to covalency between optimal doping and overdoping. Nevertheless, they are compatible with our scenario to various degrees.

A proposal has been made in which the $\pi$-bonded planar O $p_{x,y}$ orbitals, perpendicular to the usually considered ones, play a similar role as the Cu $4s$ orbital in our approach~\cite{Hirsch14}. The calculation is a good example of the kind one would like to be able to do systematically and precisely.

A large DFT calculation~\cite{Tahir-Kheli11} of a 2D slab of doped LSCO illustrates the real-space situation corresponding to the Fermi-arc regime in inverse space. It features short-range bond disorder and percolation channels.

The line T$^*$ vs. $p$ has been identified with an orbital transition~\cite{Mazumdar18}. This idea seems compatible with our observation that the hole begins to delocalize to the right of the T$^*$ line, cf.\ Fig.~\ref{fig2-transport}(c). Our proposal of an accompanying transition of the \emph{bond} from Mott-like ionicity to covalency is a separate issue, because it is possible to have a purely ionic orbital transition, without changing the nature of the bond.

Motivated by Ref.~\cite{Pavarini01}, the intraatomic Cu $4s$--$3d$ exchange $J_{sd}$ has been proposed as the Cooper-pairing mechanism~\cite{Mishonov02}. In light of the observation~\cite{Pavarini01} that the $4s$ orbital needs to be pure for a high T$_c$, we would guess that mixing by $J_{sd}$ is detrimental to SC. Intuitively we expect the $p$-part, not the $s$-part, of the $2p$--$4s$--$2p$ hopping process to scatter on the $3d_{x^2-y^2}$ orbital.

It is also possible that hopping via the $s$-complex itself provides an attraction, in the sense that the virtual intermediate state of one hole passing through the $s$-complex is attractive to another, with a small retardation. We hope to investigate this possibility in future work.

\subsection{Perspectives}

The discovery of high-T$_c$ SC has emphasized the importance of large Coulomb scales for metallic functionality. The issue is particularly sharpened in the present work, where a dual role for the O $2p$ orbitals is proposed, simultaneously partaking in the localized hole, shared with the Cu $3d$ orbital, and in the SC Fermi liquid.

The separation of these two electronic subsystems presents an ideal opportunity to address this issue further by advanced spectroscopic tools. Once the itinerant subsystem is identified as a (pseudogapped) Fermi-liquid, its responses in various observables are clear and well-defined. Thus, its contribution can be unambiguously identified and subtracted from the overall signal. The remaining part simply corresponds to the localized charge, which can thus be fully characterized. In the first place, one should carry out such an analysis on optical-conductivity and ARPES data, although all other experimental probes are of interest too.  

An early work which treats this localized/itinerant duality directly and physically is that of Kikoin and Flerov on the valence-transition problem~\cite{Kikoin79}. They introduce a canonical transformation which allows the same electron to have a Gibbs distribution on a single impurity and a Fermi distribution in the band to which it couples. This result is a concrete realization of the ionic/covalent duality, on which we rely, without resorting to artificial (slave) particles or effective baths (molecular fields, in their words). They find that the valence transition must be gradual on the average, exactly as we observe, even as it is discontinuous on each individual site, and show explicitly that discontinuities (critical behavior) in average occupation numbers are artefacts of effective-site constructions.

There is much room for progress along these lines even today. Toolboxes for systematic construction and testing of realistic, not effective, few-electron \emph{ans\"atze} for the local part of extended systems are under intense development~\cite{Lihm21}. They complement periodic DFT approaches which infer the local situation \emph{a posteriori} from calculations in the plane-wave basis. One such project, still in its infancy, has been started by one of us, in which local wave functions are generated algorithmically, based on a radical insight into the geometry of many-body Hilbert space~\cite{Sunko20}. Hopefully, some natively local approach will eventually improve both qualitative and operative understanding of the intracell electronic environment in functional materials.

\subsection{Two-dimensional superconductivity}

Finally, we recall that the observed two-dimensionality of cuprate SC~\cite{Gozar08,Wang12,Yu19} has been puzzling because of the well-known Hohenberg-Mermin-Wagner (HMW) theorem~\cite{Hohenberg67,Mermin66}, which precludes SC and magnetism in 2D above zero temperature. This theoretical worry has recently been put to rest by showing that the HMW argument, while mathematically exact, is physically ineffective~\cite{Palle21}. The infrared fluctuations which would suppress T$_c$ are simply cut off by the sample size. For SC, the HMW theorem matters only if the sample is astronomically large. In 2D, the SC T$_c$ is not affected as long as the linear sample size $L$ satisfies~\cite{Palle21}
\begin{equation}
L/a<\exp\left(T_F/T_c\right),
\end{equation}
where $a$ is the lattice constant and T$_F$ is the Fermi temperature. 

Moreover, invoking this formula to explain the observation of undiminished SC in macroscopic cuprate samples requires the Fermi energy of the SC metal to be much higher than k$_\mathrm{B}$T$_c$, thus excluding theoretical approaches which would have it low, in agreement with our observations. In other words, the failure of the HMW prediction in cuprates does permit an important physical inference about them: the band-width of the SC metal is large enough for the standard textbook picture of a Fermi edge deep inside the band to be applicable. This insight, coupled with the already cited observation~\cite{Pavarini01,Andersen01} that T$_c$ is highest when the Cu $4s$ orbital is \emph{least} hybridized with the rest of the $s$-complex perpendicular to the plane, clearly implies that SC in cuprates is natively a 2D phenomenon, becoming 3D only as a secondary effect.

\section{Conclusion}

The description of SC in cuprates in this work is based on the simplest reading of observational data and careful application of well-known concepts in physics and chemistry. The main message is that there is no need to introduce exotic new concepts once the relevant degrees of freedom, symmetries, and energy scales are prioritized in the order suggested by incontestable experimental facts.

The key new issue, probably valid across a wide range of modern functional materials, is that the functional electrons cannot be separated from the chemical background which is responsible for the binding energy. Indeed, the same oxygen $p$ orbitals have a dual role in the cuprates, partaking both in the (glue-providing) localized hole and in the (SC) Fermi liquid. Neglecting this duality by assigning Coulomb background effects to the intraorbital interactions necessarily leads to a mystification of the latter. One can by contrast only lift up the example of K.~A.~M\"uller, who has steadfastly followed the same route of physical reasoning over the past thirty-five years which led him to the discovery in the first place. His disciplined persistence in following the measurements, neither ignoring nor overtaking them, will hopefully both be remembered as the hallmark of his generation and become the lodestar for the next.

K.~A.~M\"uller has opened a new chapter in condensed-matter physics by the discovery of high-T$_c$ superconductivity in the cuprates. This discovery was seminal in two ways. First, the unprecedented interest and hopes it raised have transformed the field in terms of measurement capabilities, which are now ever more available as commodities to a steadily increasing spectrum of investigations in functional materials and thin films. Second, it has sensitized the community to the limitations of foundational theoretical approaches, which have traditionally shunned chemical details.

Largely due to the challenge of high-T$_c$ superconductivity, a new research frontier has opened at the interface of physics and chemistry, whose outlines and range are only dimly perceived even by the present generation. In this work, we have given a consistent vision of cuprate superconductivity which embodies both the specifics of their chemistry and the universality of their physics. We are honored to present it as a tribute to the founder of our research field on his 95th birthday.

\begin{acknowledgements}
We have benefited from discussions with O.~S.~Bari\v{s}i\'c, S.~Benhabib, I.~Bo\v{z}ovi\'c, T. Feh\'er, L.~Forr\'o, M. Greven, J.~E. Hirsch, S.~Mazumdar, D.~Pavuna, D. Pelc, M. Po\v{z}ek, D. Rybicki, W.~Tabi\'s, E.~Tuti\v{s}, and T.~Valla. We gratefully acknowledge our debt to the late Academician S.~Bari\v{s}i\'c.
\end{acknowledgements}

%
%

\bibliographystyle{spphys}       
\bibliography{kamueller_festschrift.bib}   

\end{document}